\documentclass[prl,aps,twocolumn]{revtex4}
\usepackage{graphicx}
\begin{document}
\title{Graphene random laser}
\author{A. Marini$^{1,}$}
\email{andrea.marini@icfo.es}
\author{F.~J.~Garc\'{\i}a~de~Abajo$^{1,2}$}
\affiliation{$^1$ICFO-Institut de Ciencies Fotoniques, The Barcelona Institute of Science and Technology, 08860 Castelldefels (Barcelona), Spain}
\affiliation{$^2$ICREA-Instituci\'o Catalana de Recerca i Estudis Avan\c{c}ats, Passeig Llu\'is Companys 23, 08010 Barcelona, Spain}
\date{\today}
\begin{abstract}
Manipulating and controlling the optical energy flow inside random media is a research frontier of photonics and the basis 
of novel laser designs. In particular, light amplification in randomly dispersed active inclusions under external pumping 
has been extensively investigated, although it still lacks external tunability, reproducibility, and control over the beam 
spatial pattern, thus hindering its application in practical devices. Here we show that a graphene random metamaterial 
provides the means to overcome these limitations through its extraordinarily-low threshold for saturable absorption. The 
nonlinear properties of nano-graphene combined with an optically pumped gain medium allow us to controllably tune the 
system from chaotic to stable single-mode lasing. Our results hold great potential for the development of single-mode 
cavity-free lasers with engineered beam patterns in disordered media.
\end{abstract}
\maketitle

\section{Introduction}

Laser operation is usually achieved through three basic elements: an amplifying medium, an external pumping setup, and an 
optical cavity that confines and shapes the emitted light in well-determined modes and directions. However, several modern 
approaches are extending this traditional laser paradigm into new avenues. For instance, the fastly developing fiber-laser 
technology replaces the optical cavity with photonic fibers, thus enabling large average powers and very high beam qualities 
\cite{Jauregui2013}. Additionally, the advent of nano-plasmonic materials has prompted the demonstration of plasmon stimulated 
emission in metallic nanoparticles \cite{Bergman2003,Noginov2009,Stockman2010} and waveguides embedding gaining media 
\cite{Noginov2008,Marini2009,Bolger2010,DeLeon2010}, which enable lasing of 
sub-wavelength beams. In a complementary direction, light amplification has been achieved in artificially engineered materials 
-- metamaterials -- that hold promise as planar sources of spatially and temporally coherent radiation 
\cite{Zheludev2008}, for compensating losses in negative-index media \cite{Wuestner2010}, and for achieving cavity-free 
lasing in the stopped-light regime \cite{Hess2012}.

Cavity-free stimulated emission of radiation has been widely studied in random lasers (RLs) 
\cite{WiersmaNatPhys2008,Gottardo2008,Wiersma2013}, where the optical cavity modes of traditional lasers are 
replaced with multiple scattering in disordered media, while the interplay between gain and scattering determines the lasing 
properties. The physics behind RLs is rich and involved, as multiple scattering of light in disordered media leads to 
very complex electromagnetic mode structures. Indeed, RL emission is typically highly multi-mode owing to the co-existence 
of narrowly confined modes ensuing from Anderson localization (AL) \cite{Sebbah2002,SegevNatPhoton2013} and extended modes
\cite{Mujumdar2004,Fallert2009}. Interestingly, light acquires a glassy behavior in RLs since nonlinear dynamics involves 
a replica-symmetry breaking phase transition \cite{Angelani2006,Ghofraniha2015}. Additionally, dispersing plasmonic nanoparticles 
in RLs enables the control of lasing resonances \cite{Meng2008} and coherent feedback \cite{Meng2008bis}. 
In spite of their striking potential applications, RLs lack external tunability, reproducibility, and control 
over the spatial pattern of the output beam. Overcoming these limitations is central for the development and application of 
cost-effective cavity-free lasers.  

Inspired by the aforementioned challenges, here we investigate the optical properties of randomly-oriented undoped graphene 
flakes embedded in externally pumped amplifying media. We demonstrate a novel mechanism leading to stable and tunable 
single-mode cavity-free lasing characterized by a well-determined and highly coherent spatial pattern. We find that the 
transverse size of the localized output beam, ranging from a few to several hundreds microns, can be accurately manipulated 
through the external pumping and through the volume density of graphene flakes. This cavity-free lasing mechanism profoundly 
relies on the extraordinary optical properties of graphene 
\cite{Bonaccorso2010,Bao2012,Javier2014}, which has already been employed in photonics for ultrafast 
photodetectors \cite{Mueller2010}, optical modulation \cite{Liu2011,Renwen2015}, 
molecular sensing \cite{Rodrigo2015,Marini2015bis}, and several nonlinear 
applications \cite{Gullans2013,DongJPB2013,Smirnova2014,Cox2014}. Graphene's salient feature underpinning cavity-free lasing 
ensues from its highly-saturated absorption at rather modest light intensities \cite{Bao2009,Xing2010}, a 
remarkable property which has already been exploited for mode-locking in ultrafast fiber-lasers \cite{Sun2010,Martinez2013}.

\section{A cavity-free graphene laser design} 

\begin{figure}[t]
\centering
\begin{center}
\includegraphics[width=0.5\textwidth]{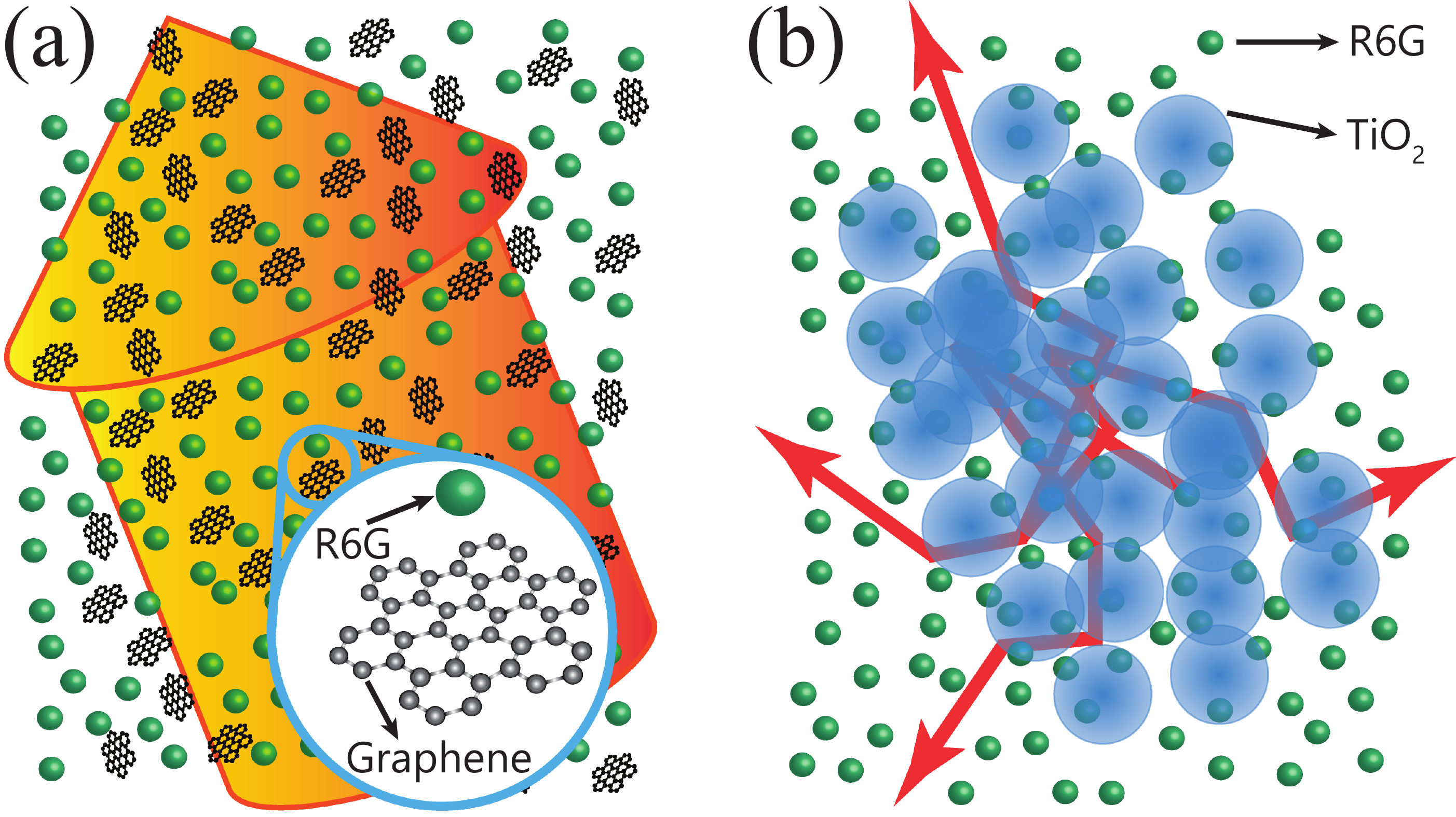}
\caption{{\bf Graphene versus traditional random lasers}. (a) Our proposed graphene cavity-free laser consists of a mixture of 
subwavelength graphene flakes and dispersed Rhodamine 6G molecules embedded in PMMA. We envision graphene flakes  of $10$'s nm 
in size, much smaller than the wavelength of light, which thus propagates in an effective medium with an exotic optical response. 
(b) In contrast, a traditional random laser uses dielectric particles instead of graphene, with sizes and separations that are 
comparable with the light wavelength in order to efficiently produce multiple scattering.}
\label{Fig1}
\end{center}
\end{figure}

For our proof-of-concept, we consider a disordered medium composed of undoped graphene flakes and Rhodamine 6G (R6G) dyes 
dispersed in polymethyl methacrylate (PMMA), as schematically depicted in Fig. \ref{Fig1}(a). Graphene is routinely exfoliated 
from graphite and dispersed in several solutions including dimethylformamide (DMF), which can be also used to dissolve R6G and 
PMMA. We remark that the practical fabrication of such disordered medium does not involve advanced nano-fabrication techniques. 
For simplicity, we assume that the graphene flakes are disks with volume density $N$ and diameter $D=30$ nm, for which quantum 
many-body and finite-size effects are negligible \cite{Cox2014}. R6G constitutes the optically amplifying medium, in which 
population inversion can be achieved through a frequency-doubled Nd-YAG laser-pump beam at wavelength $\lambda = 532$ nm. The 
aimed wavelength of laser operation is $\lambda = 593$ nm, where R6G has its peak emission. Under these conditions, graphene 
flakes are much smaller than the optical wavelength and multiple scattering, which occurs in the quasi-static regime, is 
captured by the effective permittivity of the composite material. To place this in context, typical RLs also work with R6G 
dispersed in PMMA \cite{Leonetti2013}, but in contrast, multiple scattering is efficiently achieved through randomly positioned 
particles whose dimensions are comparable with the optical wavelength [see Fig. \ref{Fig1}(b)]. Remarkably, the proposed setup 
[see Fig. \ref{Fig1}(a)] enables the engineering of the composite optical response in order to achieve stable single-mode 
operation and to manipulate the beam spatial pattern.

\section{Saturable absorption of graphene} 

\begin{figure}[t]
\centering
\begin{center}
\includegraphics[width=0.5\textwidth]{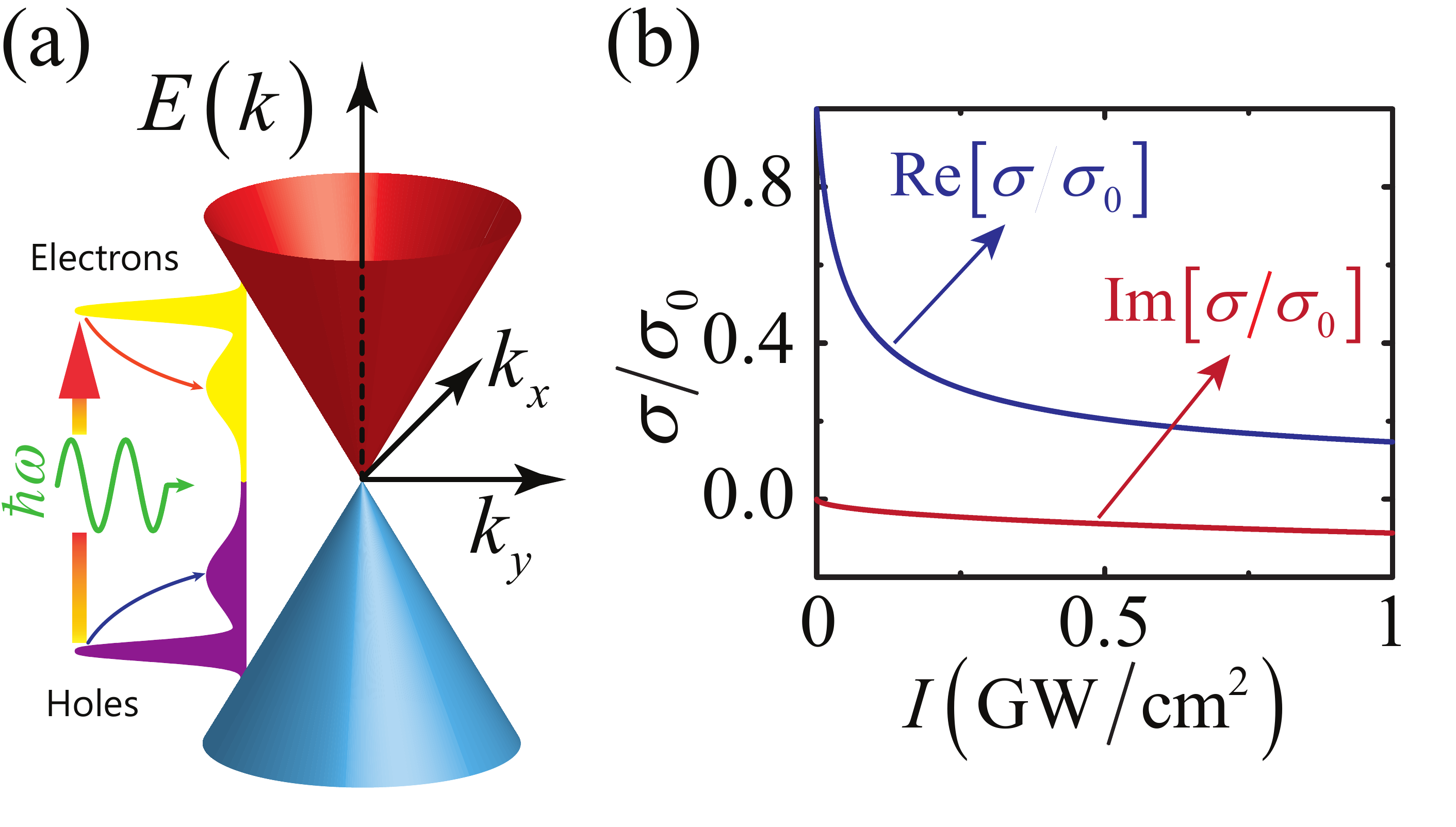}
\caption{{\bf Saturable absorption of graphene}. (a) Conical band-structure around the Dirac points. Upper 
$E_+(k) = +\hbar v_{\rm F} k$ and lower $E_-(k) = -\hbar v_{\rm F} k$ energy bands depend linearly on the electron wave-vector 
$k$. Interband transitions produced by impinging photons with energy $\hbar\omega$ lead to an out-of-equilibrium electron 
distribution, which then relaxes via electron collisions. After an initial transient time, optical pumping and electron 
relaxation compensate with each other, and absorption is saturated due to partial Pauli blocking.  (b) Intensity-dependent 
conductivity $\sigma(I)$ normalized to the universal conductivity $\sigma_0 = e^2/(4\hbar)$.}
\label{Fig2}
\end{center}
\end{figure}

The conical band structure of undoped graphene around the Dirac points [see Fig. \ref{Fig2}.(a)] is responsible for its 
unique properties. Valence electrons in this material behave as $2$-dimensional massless Dirac fermions ($2$DMDFs) with 
constant Fermi velocity $v_{\rm F} \simeq 10^6$ m$/$s \cite{Novoselov2004,Novoselov2005}. As a consequence, graphene is 
resonant over a broad range of optical wavelengths, undergoing an absorbance $\alpha_0 \approx \pi / 137$ in the limit of 
small excitation intensities. The linear optical conductivity $\sigma_0 = e^2/(4\hbar)$ is dispersionless in this optical 
regime. However, at higher light intensities, absorption saturates due to the nonlinear dynamics of graphene electrons 
[see Fig. \ref{Fig2}.(a)]. In simple terms, in the strong-field regime, electron-hole recombination produced by electron 
collisions can balance light-induced interband absorption, which is in turn partially inhibited by Pauli blocking of 
out-of-equilibrium electrons in the conduction band [see Fig. \ref{Fig2}.(a)]. In most materials, this leads to a 
light-intensity ($I$) dependence of the absorption rate $\alpha$, which typically follows a law 
$\alpha(I) = \alpha_0/[1+I/I_{\rm S}]$, where $I_{\rm S}$ is the saturation intensity. However, we find that the peculiar 
band structure of graphene produces a different intensity dependence of the absorbance: 
$\alpha(I) = \alpha_0/\sqrt{1+I/I_{\rm S}}$, where $I_{\rm S} =  137 \hbar \omega_{\rm S}^2 \omega^2 / (8 \pi v_{\rm F}^2)$
 and $\omega_{\rm S} = 6.16 $ rad ps$^{-1}$. Interestingly, we find a remarkably small value of 
$I_{\rm S} \simeq 22$ MW$/$cm$^2$ ($\lambda = 593$ nm), which is consistent with experimental findings \cite{Bao2009,Xing2010}. 
The intensity-dependent conductivity $\sigma (I)$, which is plotted in Fig. \ref{Fig2}(b), can be derived directly from 
the graphene Bloch equations (GBEs) \cite{Ishikawa2013} (see Appendix I). 

\section{ Averaged optical response} 

We model R6G amplification through the traditional Bloch description 
of two-level systems (see Appendix II), where the external optical pumping is assumed to yield a stable population inversion. 
For monochromatic waves, Bloch dynamics can be solved analytically and, at resonance, reduces to a purely imaginary susceptibility 
accounting for gain: $\chi_{\rm R6G} = - i ( g_0 / k_0) / [1 + I/I_{\rm S}^{\rm R6G}]$, where $k_0 = 2\pi / \lambda$ is the vacuum 
wave-vector, $\lambda = 593$ nm (see above), $g_0$ is the unsaturated gain coefficient (which depends on R6G density and can be 
tuned through the external pump at $\lambda = 532$ nm), and $I_{\rm S}^{\rm R6G} \simeq 150$ MW$/$cm$^2$ is the R6G saturation 
intensity \cite{Nithyaja2011}. Unsaturated gain values of about $g_0 \simeq 400$ cm$^{-1}$ have been experimentally 
demonstrated with R6G \cite{Noginov2008}. PMMA contributes to the optical response through a background 
dielectric constant $\epsilon_{\rm b} \simeq 2.23$ at the operating wavelength. Subwavelength randomly-oriented graphene disks 
are modeled through the standard electrostatic approach \cite{Javier2014}. The optical response of undoped graphene disks is 
thoroughly accounted for by the first dipolar resonance tail, which gives the polarizability 
$\alpha_{\rm G} = D^3 / \{3/(2\epsilon_{\rm b}) - 8 i \epsilon_0 c D / [\lambda \sigma ( I )]\}$. The total response of 
the system is finally calculated through the Clausius-Mossotti effective-medium theory in the limit of small graphene density 
(see Appendix III): 
$\epsilon_{\rm eff} (I) \simeq \epsilon_{\rm b} + \chi_{\rm R6G} + (2/3) N \alpha_{\rm G}$, where the factor $2/3$ accounts for 
the random orientation of the disks. The optical properties of the random medium, including PMMA, saturated 
R6G amplification, graphene absorption and the induced phase shift, are thus fully contained within $\epsilon_{\rm eff} (I)$.

\begin{figure}[t]
\centering
\begin{center}
\includegraphics[width=0.5\textwidth]{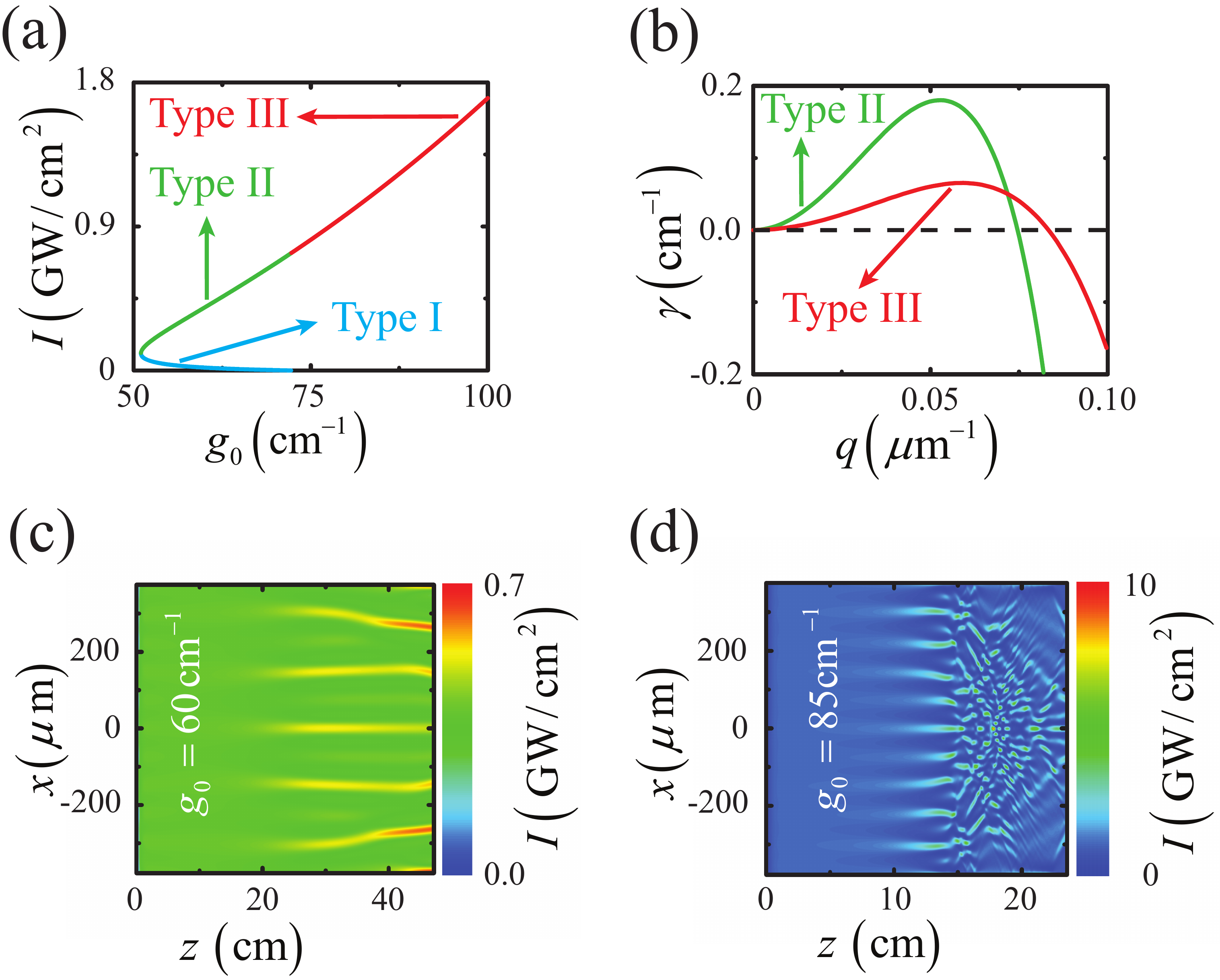}
\caption{{\bf Extended lasing modes}. (a) Existence curve of extended modes, showing their intensity 
$I = (1/2)\epsilon_0 c |A_0|^2$ against the gain coefficient $g_0$. We identify three mode types: I and II coexist in the 
sub-critical domain, while III only exists in the over-critical domain. (b) Gain spectrum depicting the instability growth 
coefficient $\gamma$ against the small-amplitude perturbation wave-vector $q$ for type II ($g_0 = 60$ cm$^{-1}$) and III 
($g_0 = 85$ cm$^{-1}$) modes; instability occurs when $\gamma > 0$. (c,d) Propagation of (c) type II and (d) type III perturbed 
extended modes for the same gain values as in (b). Graphene flakes are assumed to be disks of diameter $D = 30$ nm and density
$N = 10^3$ $\mu {\rm m}^{-3}$.}
\label{Fig3}
\end{center}
\end{figure}

\section{Dissipative optical dynamics} 

Nonlinear propagation of monochromatic beams in the effective medium 
under consideration is modeled through the slowly-varying envelope approach (see Appendix IV), where the optical field is 
expressed as ${\bf E}({\bf r},t) = A ({\bf r}_{\bot},z) e^{i k_0 (\sqrt{\epsilon_{\rm b}} z - c t)} {\bf n}$, 
${\bf r} = ({\bf r}_{\bot}, ~ z)$ is the position vector, ${\bf n}$ is a unit vector accounting for the arbitrary linear 
polarization of the beam, and the optical envelope $A (z,{\bf r}_{\bot})$ is governed by
\begin{equation}
2 i k_0 \sqrt{\epsilon_{\rm b}} \partial_z A + \nabla^2_{\bot} A + k^2 \left[\epsilon_{\rm eff} \left(|A|^2\right) - \epsilon_{\rm b}\right] A = 0. \label{PropEq}
\end{equation} 
Extended homogeneous modes of the system are calculated by setting the ansatz 
$A (z) = A_0 e^{i \beta z}$ in Eq. (\ref{PropEq}), where $\beta$ is the propagation constant correction and 
$A_0$ is the mode amplitude. We solve the ensuing nonlinear equation for $A_0$ (see Appendix IV) considering several values 
of the gain coefficent $g_0$ [see Fig. \ref{Fig3}(a)]. We find a sub-critical bifurcation of extended nonlinear modes from the 
trivial vacuum $A_0 = 0$ and identify three types of modes: Type I and II coexist in the bi-stable sub-critical domain, while 
Type III exists only in the over-critical domain [see Fig. \ref{Fig3}(a)]. We further evaluate the stability properties of these 
modes against small amplitude waves with transverse wave-vector $q$ (see Appendix IV), finding that Type I modes are always 
unstable, while Type II and III modes are unstable only for a finite range of $q$ [see Fig. \ref{Fig3}(b), where we plot the 
maximum instability growth coefficient $\gamma$ against $q$]. Besides, Type II$/$III extended modes exist on top of a 
stable/unstable background $A_0 = 0$, respectively. Thus, nonlinear dynamics in sub-critical/over-critical domains leads to 
qualitatively different phenomena [see Fig. \ref{Fig3}(c),(d)]. These results prelude the existence of localized nonlinear 
modes with complex patterns, indicating that unstable extended modes dynamically evolve into either (i) a series of stable 
filaments for sub-critical gain values [see Fig. \ref{Fig3}(c)] or (ii) a chaotic-like spatial dynamics of unstable filaments 
for over-critical gain values [see Fig. \ref{Fig3}(d)]. 

\begin{figure}[t]
\centering
\begin{center}
\includegraphics[width=0.5\textwidth]{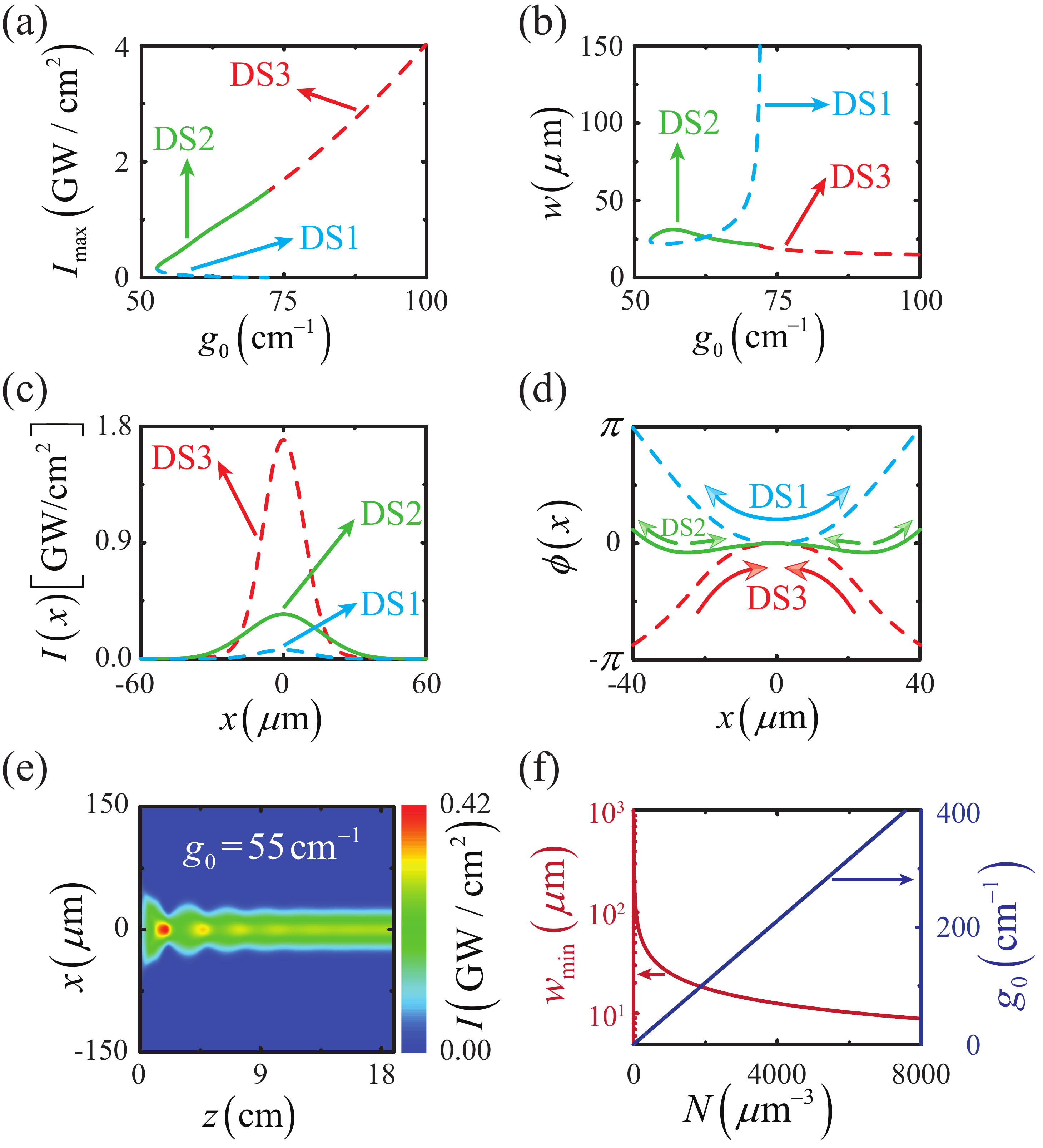}
\caption{{\bf Localized lasing modes}. (a,b) Existence curves of DSs showing (a) their maximum intensity $I_{\rm max}$ and (b) their 
width $w$ against the gain coefficient $g_0$. Three localized mode types are identified: DS1 and DS2 coexist in the sub-critical 
domain, while only DS3 exist in the over-critical domain. (c) Intensity $I(x)$ and (d) transverse phase $\phi(x)$ profiles for 
the three different localized mode types: DS1, DS2 ($g_0 = 55$ cm$^{-1}$), and DS3 ($g_0 = 75$ cm$^{-1}$). Arrows in 
(d) indicate the transverse power flow of each mode. (e) Propagation of an input Gaussian beam $I(x) = I_0e^{-(x/w)^2}$ with peak 
intensity $I_0 = 0.07$ GW$/$cm$^2$ and width $w = 70$ $\mu$m naturally evolving to the stable DS2 mode with width $w \simeq 30$ $\mu$m 
and peak intensity $I_{\rm max} \simeq 0.3$ GW$/$cm$^2$ ($g_0 = 55$ cm$^{-1}$). (f) DS2 minimum width $w_{\rm min}$ (red curve) and 
corresponding gain coefficient $g_0$ (blue curve) plotted against $N$. In every plot, graphene flakes are assumed to be disks of 
diameter $D = 30$ nm and density $N = 10^3$ $\mu {\rm m}^{-3}$ (except in (f), where $N$ is varied).}
\label{Fig4}
\end{center}
\end{figure}

We further investigate the properties of $(1+1)$D localized filaments by setting the ansatz 
$A (x,z) = A_0(x) e^{i \beta z}$ in Eq. (\ref{PropEq}) and numerically solving the ensuing nonlinear differential 
equation for $A_0(x)$ (see Appendix IV). In the context of dissipative systems, these kinds of localized nonlinear modes are 
commonly named dissipative solitons (DSs) \cite{Grelu2012}, as they involve an internal power flow enabling 
stationary propagation. We also find a sub-critical bifurcation from the trivial vacuum 
$A_0(x) = 0$ for these localized modes and we identify three types of them: DS1 and DS2 coexist in the bi-stable sub-critical domain, 
while DS3 exists only in the over-critical domain [see Fig. \ref{Fig4}(a), where we plot their maximum intensity 
$I_{\rm max} = (1/2) \epsilon_0 c ~ {\rm max} |A_0(x)|^2$ against $g_0$]. Their gain-dependent width 
$w = 2 [\int_{-\infty}^{+\infty} x^2 |A_0(x)|^2 dx / \int_{-\infty}^{+\infty} |A_0(x)|^2 dx ]^{1/2}$ is illustrated in 
Fig. \ref{Fig4}(b) for fixed graphene flake density $N = 10^3$ $\mu{\rm m}^{-3}$, and can be tuned from a few tenths to 
several hundreds of microns. Intensity profiles of DS1, DS2, and DS3 modes are plotted in Fig. \ref{Fig4}(c), while
their $x$-dependent phase $\phi = {\rm atan} [{\rm Im} A_0(x) / {\rm Re} A_0(x)]$ is shown in Fig. \ref{Fig4}(d).
Interestingly, owing to the transverse inhomogeneous phase, for every localized mode there is a peculiar internal power flow 
enabling stationary propagation within an intensity-dependent absorption/amplification environment [see Fig. \ref{Fig4}(d)]. 
Finally, we analyze mode stability properties in propagation, finding that only DS2 modes can be stable [see Fig. \ref{Fig4}(e)]. 
This leads to the conclusion that this system enables single-mode operation with spatial patterns determined by the external 
optical pump (which tunes the gain coefficient $g_0$) and the density of graphene flakes [see Fig. \ref{Fig4}(f), where we 
depict the achievable minimum DS width $w_{\rm min}$ and the corresponding gain coefficient $g_0$ against the graphene flake 
density]. Our results thus indicate that tunable single-mode cavity-free lasing can be achieved at the micrometer-scale with 
current technology. 

\section{Conclusions}

In conclusion, the extremely low-threshold saturable absorpion of graphene allows us to design an active random metamaterial 
capable of sustaining single-mode size/shape-controlled laser beams. The external pump intensity and the density of graphene 
flakes determine the regime of operation, which can be varied from chaotic to stable single-mode. Because the proposed random 
medium is a disordered mixture of currently available materials, it is promising as an inexpensive and versatile platform for 
the design of cavity-free light amplifiers and lasers.

\section{Acknowledgments}

\noindent This work has been partially supported by the European Commission (Graphene Flagship CNECT-ICT-604391 and
FP7-ICT-2013-613024-GRASP). A.M. is supported by an ICFOnest$+$ Postdoctoral Fellowship (Marie Curie COFUND program). 
A.M. acknowledges useful discussions with Joel Cox and Valerio Pruneri. 

\section{Appendix}

\subsection{I. Saturable absorption of graphene} 

In our calculations, we model light-induced out-of-equilibrium dynamics of 
$2$DMDFs through GBEs, which can be directly derived from the $2$D Dirac equation for MDFs with momentum ${\bf p}$ 
\cite{Ishikawa2013}. In this approach, MDFs are let to interact with an external monochromatic electric field 
${\bf E}(t) = {\rm Re}\left[E_0e^{- i \omega t }\right] \hat{x}$ with angular frequency $\omega$ and polarized along the 
$\hat{x}$-direction. The electromagnetic interaction is evaluated through the minimal coupling $\vec{\pi} (t) = {\bf p} + e {\bf A}(t)$, 
where the MDF quasi-momentum $\vec{\pi} (t)$ is temporally driven by the potential vector ${\bf A}(t) = - \int {\bf E}(t') dt'$.
The temporal dynamics of the ${\bf p}$-dependent coherence $\Gamma_{\bf p} (t)$ and inversion of population
$n_{\bf p} (t)$ is governed by GBEs, which in the limit $e A(t) << \hbar\omega/v_{\rm F}$ are explicitly given by 
\begin{eqnarray}
&& \dot{\Gamma}_{\bf p} = - \left( \frac{1}{T_2} + i \omega_0 \right) \Gamma_{\bf p} - \frac{ie}{2p}{\rm Re}\left[E_0e^{- i \omega t }\right] {\rm sin} \varphi n_{\bf p}, \label{GBEq1} \\
&& \dot{n}_{\bf p} = - \frac{1}{T_1}\left(n_{\bf p}+1\right) + \frac{2e}{p}{\rm Re}\left[E_0e^{- i \omega t }\right] {\rm sin} \varphi {\rm Im}\Gamma_{\bf p}, \label{GBEq2}
\end{eqnarray}
where $\omega_0 = 2 v_{\rm F} p / \hbar$, and the electron momentum is expressed in polar coordinates 
${\bf p} = p ({\rm cos}\varphi, ~ ~ {\rm sin}\varphi)$. Electron-hole recombination processes are taken into account through the
phenomelogical parameters $T_1,T_2$, which represent the population decay and the coherence dephasing time-scales, respectively.
In our calculations we assume $T_1 = T_2 = 100 {\rm fs}$. We obtain steady-state analytical solutions of Eqs. (\ref{GBEq1}-\ref{GBEq2})
by using the ansatz 
\begin{eqnarray}
&& \Gamma_{\bf p} = \Gamma_{\bf p}^+ e^{i \omega t} + \Gamma_{\bf p}^- e^{ - i \omega t}, \nonumber \\
&& n_{\bf p} = n_{\bf p}^0 + {\rm Re}\left[n_{\bf p}^{\Omega}e^{- 2 i \omega t} \right], \nonumber
\end{eqnarray}
and solving the ensuing system of algebraic equations for $\Gamma_{\bf p}^{\pm}, n_{\bf p}^0$, and $n_{\bf p}^{\Omega}$ neglecting 
third-harmonic terms. In the limit of vanishing Fermi energy and temperature, the total current density induced on the graphene is given by
\begin{eqnarray}
{\bf J} (t) = 2 e v_{\rm F} \frac{g_{\rm s}g_{\rm v}}{(2\pi\hbar)^2} \hat{x} \int_0^{2\pi} {\rm sin} \varphi d\varphi \int_0^{\infty} p \Gamma_{\bf p}(t) dp , \nonumber
\end{eqnarray}
where $g_{\rm s} = 2$ and $g_{\rm v} = 2$ are the spin and valley degeneracy factors. The graphene conductivity $\sigma$ is 
thus straightforwardly extracted from ${\bf J} = {\rm Re} [\sigma {\bf E}]$. We have numerically solved the integrals above
for several light intensities $I = (1/2) \epsilon_0 c |E_0|^2$, and found good fitting with the following analytical expression for 
$\sigma(I)$:
\begin{eqnarray}
\sigma (I) = \sigma_0 \left[ \frac{1}{ \sqrt{ 1 + I/I_{\rm S} } } - i \frac{ 1 - e^{ - \eta_1 \sqrt{ I/I_{\rm S} } } }{ \sqrt{1 + \eta_2 (I/I_{\rm S})^{0.4} } } \right], \nonumber
\end{eqnarray}
where $\sigma_0 = e^2/(4\hbar)$, $I_{\rm S} = 137 \hbar \omega_{\rm S}^2 \omega^2 / (8\pi v_{\rm F}^2)$, 
$\omega_{\rm S} = 6.16$ rad ps$^{-1}$, $\eta_1 = (\omega_\eta/\omega)$, $\eta_2 = (\omega_\eta/\omega)^{0.8}$, and
$\omega_\eta = 46.20$ rad ps$^{-1}$.

\subsection{II. Amplification of the active medium} 

For our proof-of-concept we consider externally pumped R6G as the gaining medium.
The optical response of R6G is modeled through the traditional Bloch equations (BEs) of two-level systems with resonant angular 
frequency $\omega_{\rm ba}$ and transition dipole moment $\mu_{\rm ba}$. The temporal dynamics of the density matrix coherence 
$\rho_{\rm ba} = r_{\rm ba} e^{-i\omega t}$ and population inversion $n_{\rm ba} = \rho_{\rm bb} - \rho_{\rm aa}$ under the 
external monochromatic driving field is governed by the BEs   
\begin{eqnarray}
&& \dot{r}_{\rm ba} = i(\omega - \omega_{\rm ba})r_{\rm ba} - \frac{1}{\tau_2} r_{\rm ba} - \frac{i}{\hbar} \mu_{\rm ba}E_0 n_{\rm ba}, \nonumber \\
&& \dot{n}_{\rm ba} = \frac{1}{\tau_1}\left[n_{\rm ba}^{\rm eq} - n_{\rm ba}\right] - \frac{4}{\hbar} {\rm Im} \left[ \mu_{ba} E_0 r_{\rm ba}^* \right], \nonumber
\end{eqnarray}
where $n_{\rm ba}^{\rm eq} > 0$ is the equilibrium population inversion induced by the external pump, $\tau_1$ and $\tau_2$ are
the characteristic population decay and coherence dephasing times of R6G, respectively, and we adopt the rotating wave 
approximation. Furthermore, we assume that R6G is operating at resonance (i.e., $\omega = \omega_{\rm ba}$). Stationary solutions 
of the BEs are thus directly found by setting $\dot{r}_{\rm ba} = \dot{n}_{\rm ba} = 0$. The R6G induced polarization is in turn given by  
${\bf P}(t) = N_{\rm R6G} (\mu_{\rm ba}^*\rho_{\rm ba} + \mu_{\rm ba} \rho_{\rm ba}^*)$, where $N_{\rm R6G}$ is the R6G density. Hence, the 
R6G susceptibility becomes 
\begin{eqnarray}
\chi_{\rm R6G} = - i \frac{ ( g_0 / k) }{ 1 + I/I_{\rm S}^{\rm R6G} }, \nonumber 
\end{eqnarray} 
where $I_{\rm S}^{\rm R6G} \simeq 150$ MW$/$cm$^2$ is the R6G saturation intensity \cite{Nithyaja2011}. The gain coefficient $g_0$ 
depends linearly on the equilibrium population inversion $n_{\rm ba}^{\rm eq}$, so that can be controlled through the external pump, and 
can reach values $g_0 \simeq 400$ cm$^{-1}$ \cite{Noginov2008}.

\subsection{III. Averaged optical response of the disordered mixture}

The optical response of the system depends on its three underpinning 
constituents: embedding PMMA, externally-pumped R6G, and randomly-oriented sub-wavelength graphene disks with diameter $D = 30$ nm.
PMMA contributes to the optical response with a background dielectric constant $\epsilon_{\rm b} \simeq 2.23$ 
at the operating wavelength. The polarizability of subwavelength randomly-oriented graphene disks is calculated through the standard 
electrostatic approach \cite{Javier2014}, considering only the first dipolar resonance tail:
\begin{eqnarray}
\alpha_{\rm G} = \frac{ D^3 } { 3/(2\epsilon_{\rm b}) - 8 i \epsilon_0 c D / [ \lambda \sigma ( I ) ] }. \nonumber
\end{eqnarray}
The total response of the system is thus calculated through the Clausius-Mossotti effective theory 
\begin{eqnarray}
\frac{ \epsilon_{\rm eff} - \epsilon_{\rm ext} }{ \epsilon_{\rm eff} + 2 \epsilon_{\rm ext} } = \frac{ (2/3) N\alpha_{\rm G} }{ 3 \epsilon_{\rm ext} }, \nonumber
\end{eqnarray}
where the factor $2/3$ accounts for the random orientation of graphene disks and $\epsilon_{\rm ext} = \epsilon_{\rm b} + \chi_{\rm R6G}$. In the limit of small graphene density, the Clausius-Mossotti expression reduces to 
$\epsilon_{\rm eff} (I) \simeq \epsilon_{\rm b} + \chi_{\rm R6G} + (2/3) N \alpha_{\rm G}$. 

\subsection{IV. Nonlinear dissipative dynamics}

Optical propagation of monochromatic waves inside the disordered amplifying medium 
is ruled by the double-curl macroscopic Maxwell's equations with nonlinear effective constant $\epsilon_{\rm eff} (I)$:
\begin{eqnarray}
\nabla\times\nabla\times {\bf E} = k_0^2 \epsilon_{\rm eff} (I) {\bf E}. \nonumber
\end{eqnarray}
By taking the ansatz ${\bf E}({\bf r},t) = A({\bf r}_\bot,z) e^{i k_0 (\sqrt{\epsilon_{\rm b}}z - c t)} \hat{n}$, where $\hat{n}$ is the
arbitrary polarization unit vector, and adopting the slowly varying envelope approach (SVEA) $|\partial_z A|<< k_0 \sqrt{\epsilon_{\rm b}} |A|$, 
$\nabla\cdot{\bf E}\simeq 0$, one can neglect second order derivatives of the envelope $\partial_z^2 A$ obtaining Eq. (\ref{PropEq}). The SVEA ceases 
to be valid for optical beams with size comparable to the wavelength, in which case the longitudinal component of the field becomes relevant. However, we 
never approach this limit in our calculations, and the SVEA remains fully valid for the results presented here. 

Extended homogeneous nonlinear modes are calculated by setting $A(z) = A_0 e^{i\beta z}$, where 
$\beta = (k_0/2) \left\{ {\rm Re} \left[\epsilon_{\rm eff} \left(|A_0|^2\right) \right] - \epsilon_{\rm b}\right\}$ and 
the amplitude $A_0$ is fixed by the condition ${\rm Im} [\epsilon_{\rm eff}\left(|A_0|^2\right)] = 0$, which is numerically solved through
the Newton-Raphson method. The stability of extended modes against small-amplitude perturbing waves with amplitude $\delta A$ and wave-vector 
$q$ is evaluated by setting
\begin{eqnarray}
A = \left[A_0 + \delta A_1 e^{(\gamma + i \Upsilon) z + i q x} + \delta A_2^* e^{(\gamma - i \Upsilon) z - i q x}\right]e^{i\beta z}, \nonumber
\end{eqnarray}
where $\gamma$ represents the instability growth rate and $\Upsilon$ is propagation constant shift. Inserting this expression in 
Eq. (\ref{PropEq}) and linearizing with respect to the small-amplitudes $\delta A_1, \delta A_2$, we find a linear homogeneous algebraic 
system of equations, whose complex eigenvalues $\gamma + i \Upsilon$ are calculated numerically for every wave-vector $q$. Positive/negative 
growth rates $\gamma$ indicate instability/stability against small-amplitude perturbations.

Localized nonlinear modes are calculated by setting the ansatz $A(x,z) = A_0(x) e^{i\beta z}$ in Eq. (\ref{PropEq}). The ensuing differential
equation for $A_0(x)$ is transformed into a nonlinear system of algebraic equations by discretizing the spatial variable $x = x_m$ and the 
second order derivative $\partial_x^2A(x) = [A(x_{m-1})-2A(x_m)+A(x_{m+1})]/(x_m - x_{m-1})^2$ with $m = 1,2,..,M$, and by applying homogeneous 
boundary conditions $A(x_1) = A(x_M) = 0$. This nonlinear system of algebraic equations is then numerically solved through the Newton-Raphson method.
The stability of localized modes is studied in propagation. All propagation plots have been obtained by solving Eq. (\ref{PropEq}) through the
split-step fast-Fourier-transform method embedding a fourth-order Runge Kutta routine.

\end{document}